# Near-field radiative heat transfer between graphene-covered Weyl semimetals

Yang Hu[1,2], Xiaohu Wu[1,#], Xiuquan Huang[2], Haotuo Liu[3], and Mauro Antezza[4,5,#]

[1]Shandong Institute of Advanced Technology, Jinan 250100, Shandong, P. R. China

[2]School of Power and Energy, Northwestern Polytechnical University, Xi'an 710072, Shaanxi, P. R. China

[3]Key Laboratory of Advanced Manufacturing and Intelligent Technology, Ministry of Education, Harbin University of Science and Technology, Harbin, 150080, P. R. China

[4]Laboratoire Charles Coulomb (L2C) UMR 5221 CNRS-Université de Montpellier, Montpellier F- 34095, France

[5]Institut Universitaire de France, 1 rue Descartes, Paris Cedex 05 F-75231, France

Email: xiaohu.wu@iat.cn (Xiaohu Wu); mauro.antezza@umontpellier.fr (Mauro Antezza)

**Abstract**

Polariton manipulations introduce novel approaches to modulate the near-field radiative heat transfer (NFRHT). Our theoretical investigation in this study centers on NFRHT in graphene-covered Weyl semimetals (WSMs). Our findings indicate variable heat flux enhancement or attenuation, contingent on chemical potential of graphene. Enhancement or attenuation mechanisms stem from the coupling or decoupling of surface plasmon polaritons (SPPs) in the graphene/WSM heterostructure. The graphene-covered WSM photon tunneling probabilities variation is demonstrated in detail. This research enhances our comprehension of SPPs within the graphene/WSM heterostructure and suggests methods for actively controlling NFRHT.

# 1. Introduction

In the near-field regime, where the separation between two objects is on the order of or less than the thermal radiation wavelength, evanescent wave coupling facilitates photon tunneling [1-4]. This process, known as photon tunneling, predominates over photon propagation, leading to a near-field radiative heat transfer (NFRHT) significantly exceeds the blackbody limit. The resultant substantial increase in radiative heat flux enables a plethora of applications, including thermal rectification, voltaic systems, and information processing technologies [5-7].

Weyl semimetals (WSMs) have garnered significant interest due to their unique band structures [8, 9]. The focus on NFRHT utilizing WSMs has expanded markedly in recent years [10-15]. Tang et al. demonstrated that NFRHT between two magnetic WSM slabs can be modulated by relative rotation [12]. Guo et al. both theorized and validated a radiative thermal router using magnetic WSMs [10]. Furthermore, Hu et al. designed a near-field radiative thermal diode exploiting WSM nanoparticles and a WSM substrate [13]. Graphene, an emergent two-dimensional material renowned for its superior optical characteristics, is a common choice in heterostructure systems. In recent years, combining graphene with various materials has attracted widespread attention from researchers, whether theoretical or experimental [16-19]. It supports highly confined surface plasmon polaritons (SPPs) in the infrared spectrum with lower losses than traditional plasmonic materials. The optical attributes of graphene can be dynamically adjusted by its chemical potential, allowing for tunable plasmonic resonances within the infrared spectrum. The potential for SPPs coupling/decoupling in WSM-graphene heterostructures to influence photon tunneling is an open question. Studying the NFRHT between graphene-covered WSM helps us gain a deeper understanding of the physical properties of these two materials and may provide theoretical support for developing new thermal management materials and devices.

In this work, we investigated the NFRHT between graphene-covered WSM. The two configurations when graphene covers on the front/back side are discussed. The coupling/decoupling of SPPs in WSM-based heterostructures is explored.

## 2. Modeling and calculation

In our analysis, we address the fundamental scenario for Weyl semimetals (WSMs), where the disruption of time-reversal symmetry occurs as a Dirac point bifurcates into a duo of Weyl nodes of opposing chirality. The momentum space separation of each Weyl node pair is characterized by the wavevector **b**. Materials manifesting this structure, such as $EuCd_2As_2$ (**Fig. 1**(a)), have been synthesized in recent experiments. The presence of Weyl nodes distinctly influences the electromagnetic response of these materials. The displacement electric field for WSMs is described by the equation [24]:

$$\mathbf{D} = \varepsilon_d \mathbf{E} + \frac{ie^2}{4\pi^2 \hbar \omega}\left(-2b_0 \mathbf{B} + 2\mathbf{b} \times \mathbf{E}\right), \tag{1}$$

where $e$ denotes the elementary charge, $\hbar$ denotes the reduced Planck constant. Additionally, **E** corresponds to the electric field, $\omega$ to the angular frequency, **B** to the magnetic flux density, and $\varepsilon_d$ signifies the permittivity of the associated Dirac semimetal. Dirac semimetals are typically isotropic in the wavevector space without the external magnetic fields, leading to the assumption that $\varepsilon_d$ is uniform across the permittivity tensor's diagonal. Equation (1) delineates the chiral magnetic effect and the anomalous Hall effect through its first and second terms, respectively. In our study, we focus on WSMs where the Weyl nodes are energetically equivalent (i.e., $b_0 = 0$). The vector **b**, which separates the momentum of Weyl nodes and functions akin to an internal magnetic field, is directed along the positive z-axis in our chosen coordinate system (i.e., $\mathbf{b} = b\mathbf{z}$). Given these parameters, the permittivity tensor for WSMs can be expressed as follows:

$$\boldsymbol{\varepsilon} = \begin{bmatrix} \varepsilon_d & i\varepsilon_a & 0 \\ -i\varepsilon_a & \varepsilon_d & 0 \\ 0 & 0 & \varepsilon_d \end{bmatrix}, \quad (2)$$

where

$$\varepsilon_a = \frac{be^2}{2\pi^2 \varepsilon_0 \hbar \omega}. \quad (3)$$

When $b \neq 0$, the subdiagonal component $\varepsilon_a$ is nonzero, which can lead to the breakdown of Lorentz reciprocity. The calculation of the diagonal term $\varepsilon_d$, The calculation of the diagonal term. This approach encompasses both interband and intraband transitions [25, 26].

$$\varepsilon_d = \varepsilon_b + i\frac{\sigma_W}{\varepsilon_0 \omega}, \quad (4)$$

where $\sigma_W$ is the bulk conductivity of WSM [26, 27]:

$$\sigma_W = \frac{ir_s g}{6\mathrm{Re}(\Omega)\omega}\Omega G\left(\frac{E_F \Omega}{2}\right) - \frac{r_s g}{6\pi \mathrm{Re}(\Omega)\omega}\left\{\frac{4}{\Omega}\left[1+\frac{\pi^2}{3}\left(\frac{k_B T}{E_F(T)}\right)^2\right] + 8\Omega\int_0^{\xi_c}\frac{G(E_F\xi) - G\left(\frac{E_F\Omega}{2}\right)}{\Omega^2 - 4\xi^2}\xi d\xi\right\}, \quad (5)$$

Here, $\varepsilon_b$ represents the background permittivity, and the complex frequency scaled by the chemical potential of WSM is denoted as $\Omega = \hbar(\omega + i\tau^{-1})/E_F$. $E_F$ correlates with temperature $T$, which can be deduced from charge conservation [24]:

$$E_F(T) = \frac{2^{1/3}\left[9E_F(0)^3 + \sqrt{81E_F(0)^6 + 12\pi^6 k_B^6 T^6}\right]^{2/3} - 2\pi^2 3^{1/3} k_B^2 T^2}{6^{2/3}\left[9E_F(0)^3 + \sqrt{81E_F(0)^6 + 12\pi^6 k_B^6 T^6}\right]^{1/3}}. \tag{6}$$

The specific calculation parameters of WSM are in Ref. [13].

The graphene is regarded as a layer by effective permittivity with thickness $t_G = 0.3$ nm [28]:

$$\varepsilon_G = 1 + i\frac{\sigma_G}{\varepsilon_0 \omega t_G}, \tag{7}$$

$\sigma_G$ is the conductivity of the graphene sheet, including the intraband and interband transitions contributions, respectively ($\sigma_G = \sigma_D + \sigma_I$) [21]:

$$\sigma_D = \frac{i}{\omega + i/\tau}\frac{2e^2 k_B T}{\pi \hbar^2}\ln\left[2\cosh\left(\frac{\mu}{2k_B T}\right)\right], \tag{8}$$

And

$$\sigma_I = \frac{e^2}{4\hbar}\left[G\left(\frac{\hbar\omega}{2}\right) + i\frac{4\hbar\omega}{\pi}\int_0^\infty \frac{G(\xi) - G(\hbar\omega/2)}{(\hbar\omega)^2 - 4\xi^2}d\xi\right], \tag{9}$$

Where $G(\xi) = \sinh(\xi/k_B T)/[\cosh(\mu/k_B T) + \cosh(\xi/k_B T)]$. Here, $\mu$ is the chemical potential of the graphene, $\tau$ is the relaxation time with $10^{-13}$ s, $k_B$ is the Boltzmann constant, and $T$ will be set to $T_1$ or $T_2$ in the calculation corresponding to the position of the graphene sheets. The specific calculation parameters of graphene are in Ref. [21].

The conductivities of graphene $\sigma_G$ and WSM $\sigma_W$ varying with angular frequency are shown in **Fig. 1**(b). The real part of the conductivity decreases monotonically with frequency, while the imaginary part increases and then decreases, with peaks in the infrared region. The real part of the conductivities is comparable to the imaginary in the infrared region.

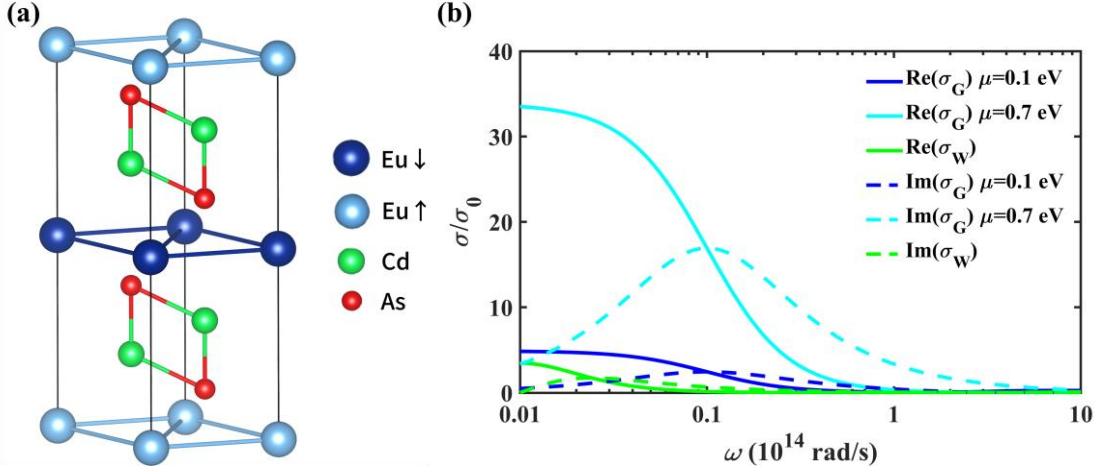

**Fig. 1** (a) Structure image of EuCd$_2$As$_2$. (b) The real (solid) and the imaginary (dash) parts of the conductivity for WSM (green) (normalized to a thickness of 0.3 nm) and for graphene (blue) at 300 K temperature in units $\sigma_0 = e^2/\hbar$ varying with the frequency.

The schematic of NFRHT between graphene-covered WSM is shown in **Fig. 2**. The NFRHT is discussed in two configurations, with graphene placed on the front and back sides of the WSM, respectively.

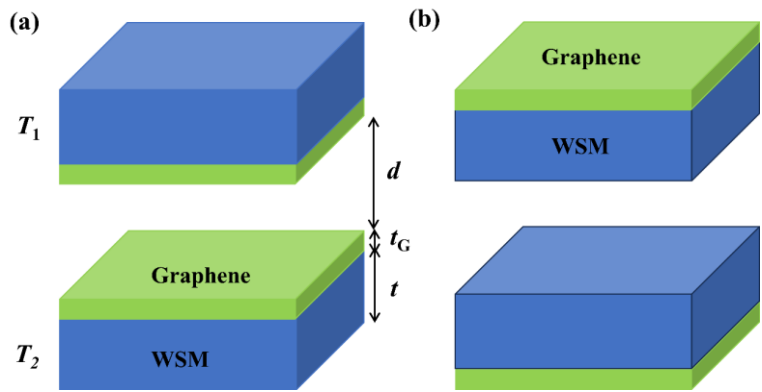

**Fig. 2** The schematic of NFRHT between graphene covered on the (a) front and (b) back side of WSM.

The NFRHT between graphene-covered WSM is calculated by [29, 30]:

$$Q = \frac{1}{8\pi^3} \int_0^\infty [\Theta(\omega,T_1) - \Theta(\omega,T_2)] d\omega \int_0^{2\pi} \int_0^\infty \xi(\omega,\beta,\phi) \beta d\beta d\phi, \qquad (10)$$

where 1/2 represent the emitter/receiver, and $\xi(\omega,\beta,\phi)$ is the photon tunneling probability, which is given by:

$$\xi(\omega,\beta,\phi) = \begin{cases} \text{Tr}\left[(\mathbf{I} - \mathbf{R}_2^*\mathbf{R}_2 - \mathbf{T}_2^*\mathbf{T}_2)\mathbf{D}(\mathbf{I} - \mathbf{R}_1^*\mathbf{R}_1 - \mathbf{T}_1^*\mathbf{T}_1)\mathbf{D}^*\right], \beta < k_0 \\ \text{Tr}\left[(\mathbf{R}_2^* - \mathbf{R}_2)\mathbf{D}(\mathbf{R}_1 - \mathbf{R}_1^*)\mathbf{D}^*\right]e^{-2|k_z|d}, \beta > k_0 \end{cases}, \qquad (11)$$

Where $k_z = \sqrt{k_0^2 - \beta^2}$, **R** and **T** is reflection/transmission coefficients tensor, $\mathbf{D} = (\mathbf{I} - \mathbf{R}_1\mathbf{R}_2 e^{2jk_z d})^{-1}$.

## 3. Results and discussion

The heat flux dependent on the chemical potential of graphene and the thickness of the WSM when graphene is covered on the front side of the WSM, as shown in **Fig. 3**(a). The colorbar takes logarithms here. The heat flux is more significant at small thicknesses of WSM and the chemical potential of graphene. The heat flux decays rapidly with increasing thickness and chemical potential. The blue lines represent the variation in heat flux between bare (WSM), as influenced by their thickness. The flux shows a monotonic decrease as the WSM thickness increases. Similarly, the green lines depict the heat flux between bare graphene sheets, which changes in response to the chemical potential of the graphene. For graphene sheets, this flux initially increases and subsequently decreases as chemical potential rises, peaking at approximately 0.1 eV. The heat flux ratio of graphene covered with WSM to bare WSM is shown in **Fig. 3**(b). After covering graphene on WSM, the ratio will be greater than 1 at smaller thicknesses ($t < 50$ nm) and smaller conductivities ($\mu < 0.15$ eV). When the configuration is graphene covered on the back side of the WSM, the same will be found with high heat flux at small thicknesses and small conductivities, as shown in **Fig. 3**(c). The heat flux ratio is shown in **Fig. 3**(d). The presence of the graphene plasmonic cavity enhances the heat flux between WSMs over a wide range of thicknesses with conductivities less than 0.2 eV.

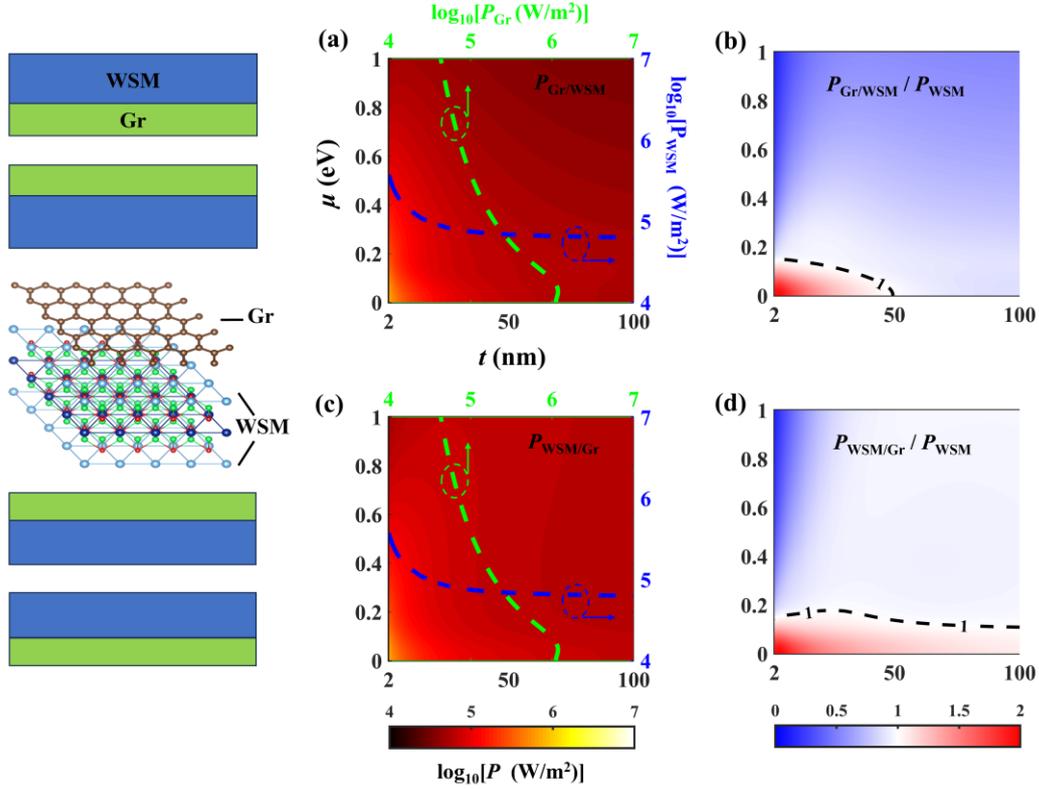

**Fig. 3** (a), (c) The near-field heat flux varying with the chemical potential of graphene and the thickness of the WSM. (b), (d) The heat flux ratio of graphene-covered WSM to bare WSM. (a), (b) Graphene covered on the front side of the WSM. (c), (d) Graphene covered on the back side of the WSM. The inset is the atom structure of graphene-WSM systems.

The spectral heat fluxes varying with angular frequency are shown in **Fig. 4**. The spectral heat fluxes distribution of graphene-covered WSM when the chemical potential of graphene is 0.1 eV are very similar for both configurations. Despite a slight decrease in amplitude, they exhibit broader resonant at around $1.75 \times 10^{14}$ rad/s than bare WSM. When the chemical potential is 0.7 eV, there are two resonance peaks in the spectrum at around $2.18\times10^{14}$ rad/s and $3.06\times10^{14}$ rad/s with significant attenuation. The amplitude of spectral heat flux when graphene-covered on the back side is much larger than the front side at the resonance peak of $2.18\times10^{14}$ rad/s, while the amplitudes $3.06\times10^{14}$ rad/s are highly consistent.

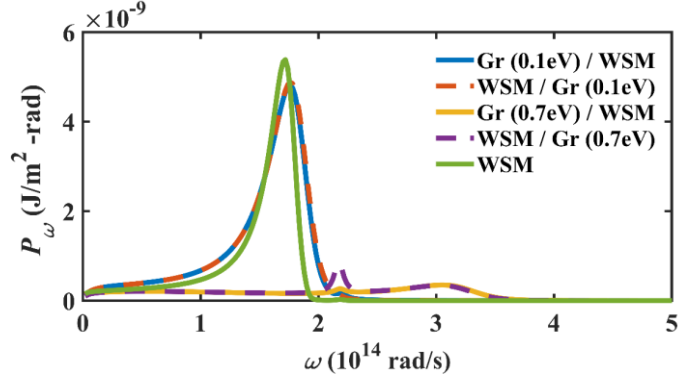

**Fig. 4** Spectral heat flux distribution in different configurations.

To explore the mechanism for the enhanced and attenuated heat flux, the photon tunneling probabilities between the emitter and the receiver distribution in frequency-wavevector space in different configurations are shown in **Fig. 5**. It is worth noting that the WSM can be regarded as isotropic in wavevector space when the momentum-separation **b** is along the *z*-direction. (By rotating the dielectric function matrix in *x-y* space, i.e., after multiplying the rotated matrix **T** and **T**$^{-1}$, the form of the matrix $\varepsilon^T$ remains unchanged. Therefore, it can be considered isotropic in the $k_x$-$k_y$ space, and we analyze the photon tunneling probabilities in frequency-wavevector space).

The bright bands depicted in **Fig. 5** represent photon tunneling probability, resulting from the excitation of SPP at various conditions. These correspond to the dispersion relationship where the denominator of Eq. (11) close to zero. When the thickness of bare WSM is 5 nm, the photon tunneling probability distribution properties (**Fig. 5**(a)) are similar to bare graphene when the chemical potential is 0.1 eV (**Fig. 5**(b)). In Fig. 5(a)/(b), the two bands represent the branches of the coupled SPPs between two WSM/graphene sheets: the lower angular frequency band is symmetric, while the higher angular frequency band is asymmetric. WSM has slimmer branches compared to graphene, indicating poorer heat transfer performance. When covering the graphene on the front/back side of WSM, the photon tunneling probabilities distribution are shown in **Figs. 5**(c) and (d). The photon tunneling probabilities are enhanced in the smaller wavevector space. When the chemical potential is 0.7 eV, a higher chemical potential can push the SPP to larger frequencies and small wavevectors, as shown in **Fig. 5**(f). At this point, the resonance of the WSM is mismatched with that of graphene, leading to the attenuation of the heat flux when covering the graphene on the WSM. When graphene is covered on the front side of WSM (**Fig. 5**(g)), the wavevector region where SPPs excitation shrinks dramatically. The resonance can be observed at around $2.2\times10^{14}$ rad/s, attributed to the decoupling between SPP in graphene and WSM. When graphene is covered on the back side of WSM (**Fig. 5**(h)), a graphene plasmonic cavity is formed in the heterogeneous system. The SPPs of WSM are excited mainly in the symmetric (lower frequencies) branch and extend toward higher wavevectors than graphene covered on the front side, leading to higher spectral heat flux.

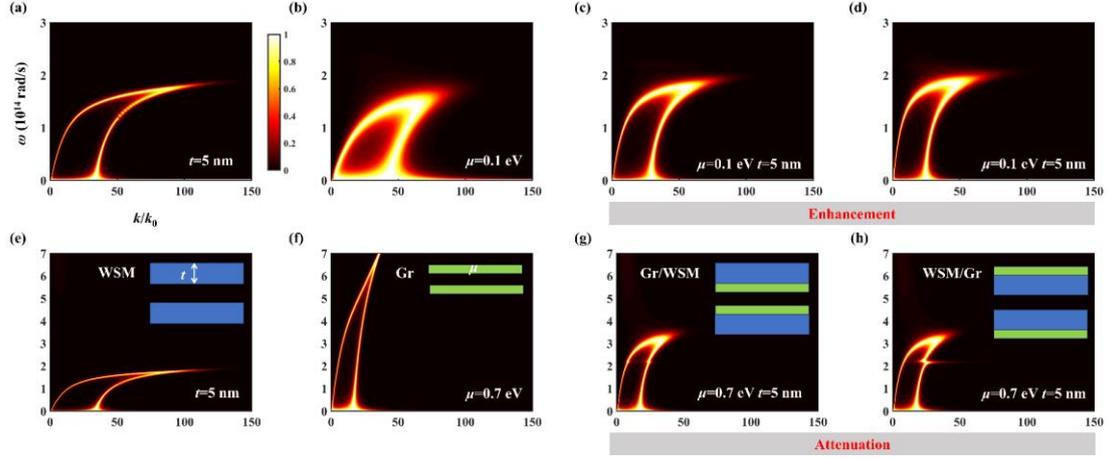

**Fig. 5** Photon tunneling probability between the emitter and the receiver varying with angular frequency and dimensionless wavevector in different configurations. (a), (e) bare WSM, (b), (f) bare graphene, (c), (g) graphene covers on the front side of the WSM, (d), (h) graphene covers on the back side of the WSM. $\mu = 0.1$ eV in (b)-(d) and $\mu = 0.7$ eV in (f)-(h).

## 4. Conclusions

This study primarily investigates NFRHT in graphene-covered WSM. It specifically examines the effects of graphene layers on both the front and back sides of WSM. Numerical analyses reveal that the interplay between SPPs in graphene and WSM can modulate the heat flux, influenced by chemical potential of graphene. Additionally, the study explores the excitation characteristics of SPPs across various heterostructure configurations. Consequently, this research enhances our understanding of polariton coupling in graphene-WSM systems and suggests methods for active control over photon tunneling in graphene-covered topological materials.

**Data Availability**

Data from this study can be provided upon reasonable request.

**Acknowledgments**

This work is supported by the National Natural Science Foundation of China (52106099), the Shandong Provincial Natural Science Foundation (ZR2022YQ57), and Taishan Scholars Program.